\documentclass[aps, prb, twocolumn, superscriptaddress]{revtex4-2} 
\usepackage{amsmath, amssymb}   
\usepackage{graphicx}           
\usepackage{dcolumn}            
\usepackage{bm}                 
\usepackage[colorlinks=true, citecolor=red, linkcolor=red, urlcolor=red]{hyperref}  
\usepackage{braket}             
\usepackage{natbib}             
\def \SM {\textcolor{black}{{supplemental material}}}

\begin{document}
	
\title{Emergent quasi-particles of spin-$1$ trimer chain}

\begin{abstract}
The recent experimental realization of emergent quasi-particles, such as spinons, doublons, and quartons, in a spin-$1/2$ trimer chain has spurred new interest in low dimensional magnetic systems. In this study, we investigate the dynamical properties of the isotropic spin-$1$ trimer chain with antiferromagnetic intra-trimer ($J>0$) and inter-trimer ($J' > 0$) exchange couplings, unveiling various quasi-particles: magnons, singletons, triplons, pentons, and heptons. For weak inter-trimer exchange coupling $J'/J \ll 1$, it  behaves as an effective spin-$1$ chain with valence bond solid (VBS) ground state. Employing density matrix renormalization group (DMRG) techniques, we compute the dynamical structure factor (DSF) which reveals a gapped magnon band along with weakly dispersive singleton, excited triplon, and penton energy bands. The evolution of these excitations with inter-trimer coupling $J'$ is also examined, providing insight into the underlying excitation mechanisms. For spin-$1$ chain, these exotic quasi-particles eventually reduce to conventional magnon excitations as $J'/J \rightarrow 1$. Our results also shed light on the rich and complex excitation spectrum of spin-$1$ trimer chains and offer unique perspectives on the dynamics in quantum spin systems.  
\end{abstract}

\author{Manodip Routh}
\thanks{The authors contributed equally to this work.}
\affiliation{Department of Condensed Matter and Materials Physics,
S. N. Bose National Centre for Basic Sciences, Kolkata 700106, India}
\author{Anutosh Biswas}
\thanks{The authors contributed equally to this work.}
\affiliation{Department of Condensed Matter and Materials Physics,
S. N. Bose National Centre for Basic Sciences, Kolkata 700106, India}
\author{Manoranjan Kumar}
\email[Corresponding author: ]{manoranjan.kumar@bose.res.in}
\affiliation{Department of Condensed Matter and Materials Physics,
S. N. Bose National Centre for Basic Sciences, Kolkata 700106, India}

\date{\today}

\maketitle	

{\it Introduction.}--- The emergence of excitons and exotic quantum phases due to quantum fluctuations and the effect of confinement of quasi-particles in one-dimensional (1D) correlated quantum systems remains a fundamental topic of research in condensed matter physics \cite{Lacroix2011}. In 1931, Bethe stated an ansatz to predict the ground state (gs) of the 1D spin-1/2 Heisenberg antiferromagnetic (HAF) model \cite{Bethe1931}. Later its spinon excitations were predicted \cite{Muller1981} and observed \cite{spin-1/2excitation}. Further extension of this ansatz successfully predict the gs as well as excitations of the generalized 1D anisotropic XXZ antiferromagnetic (AFM) model, many-body Bethe string states \cite{Bera2017, Wang2018, Bera2020, Lake2005, Lake2010, Caux2005}. 
Haldane predicted that integer and half-integer HAF spin chains are fundamentally different, with the former exhibiting a gapped spectrum and the latter a gapless spectrum  \cite{Haldane1983, Steven1993}. The gs of spin-$1$ HAF chain is a topological valence bond solid (VBS) ~\cite{Affleck1987,schollwock2002} with a gapped spectrum \cite{Haldane1983, Haldanesigma1983, Affleck1987, Steven1993, Haldane2017}, whereas the spin-$1/2$ HAF chain exhibits a spin-liquid gs with quasi-long-range order and a gapless spinon continuum \cite{Muller1981, Jacques1962, Mourigal2013}. These spin$-1$ chains are also notable as prototypes for considering topologically ordered physics \cite{Wen2017}, having a hidden nonlocal order parameter \cite{Nijs1989}.

Interestingly, the spin-$1/2$ HAF chain develop a gap due to dimerization in the chain~\cite{Bray1975,Jacobs1976,PhysRevB.19.402, PhysRevB.20.4606}, whereas weakly coupled spin-$1/2$ trimer chains remain gapless. Recently, antiferromagnetically weakly coupled spin-$1/2$ trimer chains have been shown to exhibit emergent composite excitations of novel quasi-particles, namely doublons and quartons, in addition to low-energy fractional spinon excitations \cite{Cheng2022, Bera2022, Cheng2024, Li2025, prabhakar2024}. These quasi-particles have been experimentally observed in A$_2$Cu$_3$Ge$_4$O$_{12}$ [where A $=$ Na, K] \cite{Bera2022, prabhakar2024, Cheng2024,  Stoll2018}. The effective spin-$1/2$ state on each trimer may be considered a potential candidate for qubits. The emergent composite quasi-particles exist only in the weak inter-trimer exchange limit ($J'/J \ll 1$), where $J$ and $J'$ denote the inter-trimer and intra-trimer exchanges, respectively. As $J'/J \rightarrow 1$, these states fractionalize to form the conventional spinon continuum.

In case of regular HAF spin-$1$ chain, the lowest excitation is gapped due to the formation of a massive lowest-energy triplet bound state above the valence bond solid ground state \cite{Haldane1983, Haldanesigma1983}. This excitation gap, known as the Haldane gap, is proportional to the AFM exchange interaction $J$ \cite{Haldane1983, Haldanesigma1983, Steven1993}. Field-theoretical studies of the nonlinear sigma model (NL$\sigma$M) show that the lowest excitations form a massive magnon triplet. The excitation spectrum exhibits a gap of $\Delta_H \approx 0.41J$ at wave vector $q = \pi$, and the dispersion of low-lying excitations for $q$ close to $\pi$ follows the relativistic form $\epsilon(q) = \sqrt{\Delta_{H}^2 + v^2 (q - \pi)^2}$, where the spin wave velocity is $v \approx 2.46J$. Now the question arises what happens to the excitation spectrum in a spin-$1$ trimer chain? To date, no antiferromagnetic spin-$1$ trimer compound has been reported. However, recently synthesized compounds $\text{ANi}_{3}\text{P}_{4}\text{O}_{14}$ [A $=$ Ca, Sr, Pb, Ba] consist of ferromagnetically coupled spin-$1$ trimers \cite{Masashi2012, Masashi2006, Majumder2015, Yusuf2016, Anup2018}.  

In this letter, we explore the dynamical properties of the isotropic spin-1 trimer chain with antiferromagnetic exchange couplings ($J, J' > 0$), uncovering a spectrum of novel quasi-particle excitations beyond conventional magnon modes. In the weak inter-trimer exchange limit ($J'/J \ll 1$), perturbation theory is employed to explore the effective low-energy spectrum of the system. We use correction vector method with density matrix renormalization group (DMRG) technique to compute the dynamic structure factor , revealing various excitations, including magnons, singletons, triplons, pentons, and heptons, each exhibiting unique dispersion characteristics.

\begin{figure}[h]
    \centering
    \includegraphics[width=1.0\linewidth]{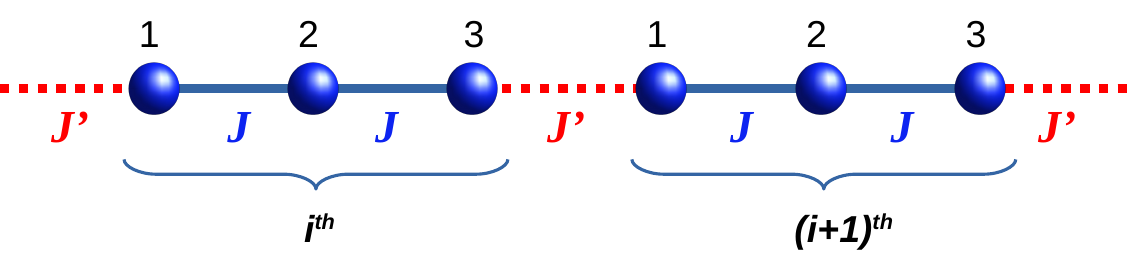}
    \caption{Schematic diagram of a 1D isotropic spin-$1$ trimer chain, where $J$ and $J'$ represent the antiferromagnetic (AFM) intra-trimer and inter-trimer exchange interactions, respectively. The indices $1$, $2$, and $3$ label the three spins within a unit cell.}
    \label{fig:spin_model}
\end{figure}

{\it Model Hamiltonian and Methods.}--- We consider a 1D isotropic spin-$1$ trimer chain, where three consecutive spins along the chain are coupled by strong exchange interactions, forming a trimer unit. These trimer units are then connected to each other by comparatively weaker interactions as shown in Fig.~\ref{fig:spin_model}. The Hamiltonian of the spin-$1$ HAF trimer model, consisting of $N/3$ trimers or $N$ spins can be expressed as follows:  
\begin{eqnarray}
	\mathcal{H} = \sum_{i=1}^{N/3} \left[ J \left(\vec{S}_{i,1} \cdot \vec{S}_{i,2} + \vec{S}_{i,2} \cdot \vec{S}_{i,3} \right) +  J' \vec{S}_{i,3} \cdot \vec{S}_{i+1,1} \right], \nonumber \\
    \label{Eq:Hamiltonian}
\end{eqnarray}
where $\vec{S}_{i,r}$ is a spin-$1$ operator acting on the $r$-th site of the $i$-th trimer. The parameters $J$ and $J'$ represent the intra-trimer and inter-trimer antiferromagnetic exchange interactions, respectively. The intra-trimer exchange interaction is set as $J = 1$, which serves as the unit of energy. Our study is restricted to the parameter range $0 \leq J'/J \leq 1$.  

We use perturbation theory and density matrix renormalization group (DMRG) method \cite{white1992density, white1993density, Dey_Maiti_Kumar_2016} to study the model Hamiltonian with periodic and open boundary conditions both.  In our DMRG calculations, we consider trimer chain with up to $N=96$ spins and retain up to $1000$ density-matrix eigenstates $m$ during the renormalization process. Retaining the large number of $m$ ensures the high accuracy of our results, with the largest discarded weight being maintained at an impressively low level of approximately $10^{-10}$.

The dynamical spin structure factor (DSF) is defined as:
\begin{equation}
S(q,\omega) = \sum_{n} \frac {\left| \langle \psi_n | S_q^{\alpha} | \psi_0 \rangle \right|^2}{(E_n - (E_0 + \omega) + i\eta)},
\end{equation}
where, $E_0$ and $E_n$ are the energies of the gs and the $n^{th}$ excited state, respectively. The $\omega$, $\eta$ and $q$ represent the energy, broadening factor and momentum, respectively. $|\psi_0\rangle$ is the gs wavefunction and $|\psi_n\rangle$ is the $n^{th}$ excited state wavefunction. If $\alpha$ denotes the $x$, $y$, and $z$ components of spin, we can define $S_q^{\alpha} = \sqrt{\frac{2\pi}{N}} \sum_{j} e^{iqj} S_j^{\alpha}$. 

The DSF of the system is calculated using the combination of the DMRG and the correction vector method which is a  well-established numerical approach \cite{Soos1989, Ramasesha1995, Ramasesha1997, Till1999, Jeckelmann2002, Nocera2016}. In this work, we use both conjugate gradient \cite{Soos1989, Ramasesha1995, Ramasesha1997, Till1999} and Krylov space approach \cite{Nocera2016}. However, all the simulations here are done using Krylov space approach. The details of the method is given in the supplemental material \cite{suppmat}.



{\it Isolated spin-$1$ trimer.}--- We begin by examining the energy spectrum and eigenfunctions of an isolated spin-$
1$ system, with the corresponding model Hamiltonian given by:
\begin{equation}
    \mathcal{H}=J(\vec{S}_1.\vec{S}_2+\vec{S}_2.\vec{S}_3)
    \label{Eq:isolated}
\end{equation}
The Hamiltonian can  be written in terms of total spins $\vec{S}_{123}=(\vec{S}_1+\vec{S}_2+\vec{S}_3)$, $\vec{S}_{13}=(\vec{S}_1+\vec{S}_3)$ and $\vec{S}_{2}$ as
\begin{align}   
\mathcal{H}(\vec{S}_{123}, \vec{S}_{13}, \vec{S}_{2}) =\frac{J}{2}(\vec{S}_{123}^{2}-\vec{S}_{13}^{2}-\vec{S}_{2}^{2})
 \label{eq:eq_ham}
\end{align} 
The details of the calculations is given in the \SM ~\cite{suppmat} and  the  energies of the spin-1 trimer can be written as,
\begin{align}
  E(S_{123}, S_{13}, S_{2}) =&\frac{J}{2}[S_{123}(S_{123}+1)-S_{13}(S_{13}+1) \nonumber\\
  &-S_{2}(S_{2}+1)].
 \label{eq:eq_ene}  
\end{align}

Each spin-$1$ site contributes three degrees of freedom; therefore, the total number of degrees of freedom in the trimer is $27$. The total spin ${S}_{123}$ of each trimer can take values $0$, $1$, $2$, and $3$. The Hamiltonian in Eq.~\eqref{eq:eq_ham} can be solved exactly, and the energies in Eq.~\eqref{eq:eq_ene} can be written in terms of $S_{123}$, $S_{13}$, and $S_2$. The various combinations of spin values $S_{123}$, $S_{13}$, and $S_2=1$ give rise to $27$ possible energy levels. The spin values, energy levels, and their degeneracies are shown in Table S1 in \SM 
~\cite{suppmat}.
 
\begin{figure*}[t]
    \centering
    \includegraphics[width=0.75\linewidth]{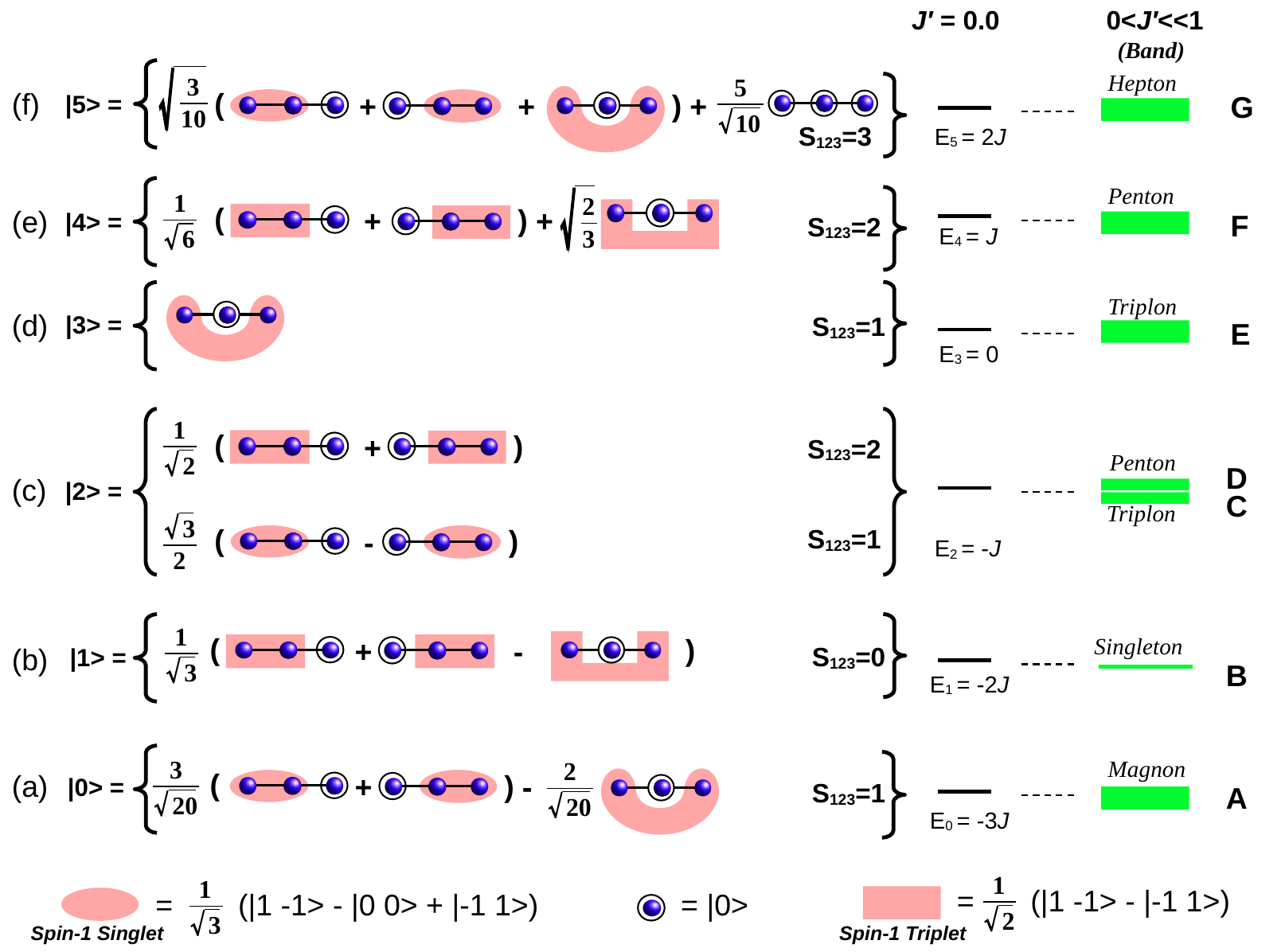}
    \caption{Energy levels of an isolated spin-$1$ trimer chain are shown in $S^{z}_{123}=0$ sector and the spin configurations are represented in terms of singlet and triplet states of two spin-$1$ and a free spin-$1$. Spin-1 singlet and triplet are represented by pink ellipse and rectangle, respectively. For weak inter-trimer coupling ($J'/J \ll 1$), the energy levels form bands of quasi-particles, magnon, singleton, triplon, penton, hepton. First column lists the state numbers, corresponding local state numbers, and their associated wavefunctions. Second column shows the discrete energy levels of an isolated spin-1 trimer($J'/J=0$). Third column presents the energy bands that emerge when inter-trimer exchange coupling $J'/J$ is introduced in the perturbative limit.}
    \label{fig:ene_spec}
\end{figure*}

Fig.~\ref{fig:ene_spec} shows the wavefunctions of all possible states in $S^{z}_{123}=0$ sector. The wavefunctions are expressed as a combination of spin-$1$ singlet, triplet and a free spin with $S^z=0$ component. However, all the wavefunctions are explicitly given in the \SM~\cite{suppmat}. The gs of an isolated spin-$1$ trimer in Eq.~\eqref{eq:eq_ham} is a threefold degenerate triplet state with energy $E_{0}(S_{123}=1) = -3J$, composed of $S_{13}=2$ and $S_{2}=1$. 
The gs in $S^z = 0$ sector is represented as a linear combination of a spin-$1$ singlet and a free spin-$1$, as shown in Fig.~\ref{fig:ene_spec}(a). The free edge spin-$1$ configurations are more prominent compared to the free spin-$1$ in the middle of the trimer. 

The lowest excitation is a unique singlet state at $E_{1}(S_{123}=0)=-2J$, consisting of $S_{13}=1$ and $S_2=1$, we call the quasi-particle as {\it singleton}. The spin configuration of the state is shown in Fig.~\ref{fig:ene_spec}(b). This state is represented as a combination of spin-$1$ triplet and a free spin-$1$, leading to an effective spin $S_{123}=0$. There are three possible arrangements of spin-$1$ triplet and the free spin-$1$.

The second excited state is eightfold degenerate with energy $E_{2}(S_{123}=1,S_{123}=2)=-J$ due to the formation of two spin states: a triplet with $S_{123}=1$ ($S_{13}=1$, $S_{2}=1$) and a pentate with $S_{123}=2$ ($S_{13}=2$, $S_{2}=1$), as shown in Fig.~\ref{fig:ene_spec}(c). These excitations are called \textit{triplon} and \textit{penton}, respectively. For $S_{123}=1$, the state is threefold degenerate and can be represented as a product state of a spin-$1$ singlet and a free spin-$1$ at the edge and have two possibilities of the spin-$1$ singlet and the free spin-$1$, as shown in Fig.~\ref{fig:ene_spec}(c). Similarly, for pentate state it is product of edge spin and its two possibilities.


The third excited state is also a threefold degenerate triplet excited state with energy $E_{3}(S_{123}=1)=0$. This state is composed of $S_{2}=1$ and $S_{13}=0$, 
where the first and third spin-$1$ form a singlet state, while the middle spin remains a free spin, as shown in Fig.~\ref{fig:ene_spec}(d) in the $S^z=0$ sector.

The fourth excited state is fivefold degenerate with energy $E_4(S_{123}=2)=J$ and is composed of $S_{13}=1$ and $S_2=1$. We refer to the corresponding quasi-particle as a \textit{penton}. The pictorial representation of this state for $S_{123}^z=0$ sector is shown in Fig.~\ref{fig:ene_spec}(e). This state is represented as a linear combination of spin-$1$ triplet and a free spin-$1$, leading to an effective spin $S_{123}=2$.


The highest excited energy state is sevenfold degenerate at energy $E_{5}(S_{123}=3)=2J$, where all three spins combine to form $S_{123}=3$ ($S_{13}=2$, $S_{2}=1$). This state is represented pictorially in Fig.~\ref{fig:ene_spec}(f). The corresponding excited quasi-particle has seven degrees of freedom and is referred to as a \textit{hepton}.


{\it Coupled trimers.}--- In the decoupled trimer limit, the gs of each trimer has total spin $S_{123} = 1$ and exhibits threefold degeneracy. Therefore, each trimer can be treated as a single effective site with a effective spin-$1$ degree of freedom, represented by the spin operator $\vec{\tau}$.
In the weak inter-trimer exchange limit, $J'/J \ll 1$, the system can be effectively described as an isotropic Heisenberg antiferromagnetic (HAF) chain with effective spin $\tau = 1$ at each site, where each site corresponds to a trimer. The effective Hamiltonian of the chain, $\mathcal{H}_{\text{eff}}$, is obtained through first-order degenerate perturbation theory and expressed as:

\begin{equation}
    \mathcal{H}_{eff}=J_{eff}\sum_{i}\vec{\tau}_{i}.\vec{\tau}_{i+1} + O^n(J'),
\end{equation}
where $\vec{\tau}_i$ represents the effective spin-$1$ at $i-$th site and $J_{eff}=\frac{9J'}{16J}$. The behavior of effective spin $\tau=1$ chain should be similar to the spin-$1$ chain. The details of the calculations are provided in \SM
~\cite{suppmat}.   

\begin{figure}[t]
    \centering
    \includegraphics[width=1.0\linewidth]{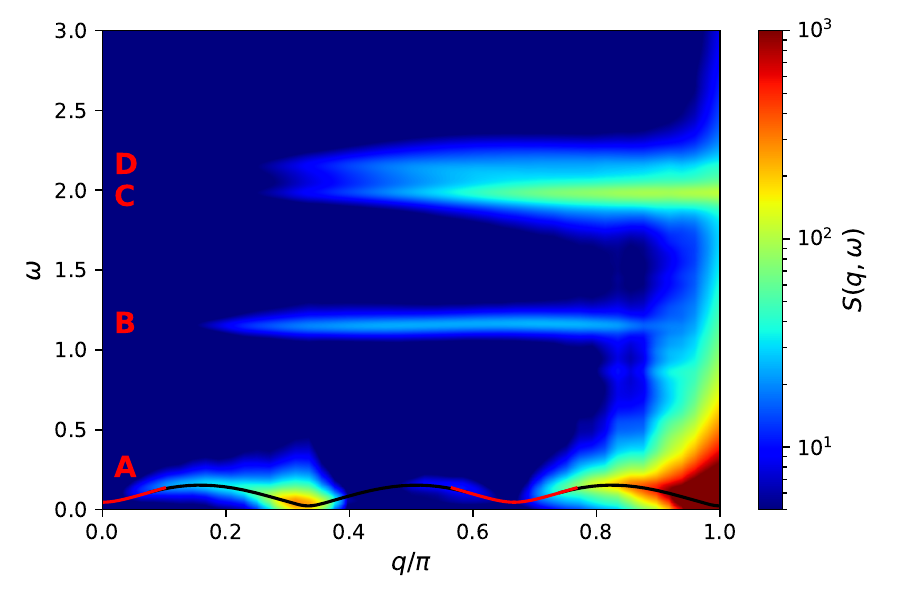}
    \caption{Dynamic spin structure factor $S(q, \omega)$, computed using DMRG for a trimer chain of system size $N=96$ at weak inter-trimer coupling $J'/J = 0.1$, with a broadening factor of $\eta = 0.05$. The magnon band is labeled as A, while B, C, and D correspond to singleton, triplon, and penton excitations, respectively. The magnon band in the reduced Brillouin zone (BZ) is fitted using the expression $\omega (q) $. The black line represents the single-magnon dispersion, while the red line shows the two-magnon dispersion given by $2\omega(\pi-q/2)$.}
    \label{fig:DSF}
\end{figure}

The trimer chain exhibits translational invariance with a unit cell consisting of three spins and maintains rotational symmetry. Consequently, the effective Hamiltonian must preserve both translation symmetry and SU(2) symmetry.
For small $J'/J$, the internal excitations of the trimers give rise to weakly dispersive bands. Fig.~\ref{fig:DSF} shows the DSF for $J'/J=0.1$. The lowest band is a single magnon band A, governed by the effective Hamiltonian $\mathcal{H}_{eff}$ of effective spin $\tau=1$. The higher energy bands are weakly dispersive singleton, triplon and penton denoted by energy band B, C, and D in Fig.~\ref{fig:DSF} indicate certain characteristics of a single trimer excitation persist in the weak coupling regime of $J'/J$.

{\it Energy band A.}--- The lowest dispersive energy band, labeled as the $A$ band in the DSF, $S(q,\omega)$ corresponds to magnon band within a reduced Brillouin zone. The full Brillouin zone (BZ), spanning $(0, 2\pi)$, is reduced to the interval $(0, 2\pi/3)$ due to the unit cell, which comprises three sites per trimer. Consequently, the lowest dispersive single-magnon band, originally extending over $(0, \pi)$ at $J'/J=1$, fragments into three distinct regions: $q \in [0, \pi/3]$, $q \in [\pi/3, 2\pi/3]$, and $q \in [2\pi/3, \pi]$. This lowest energy band exhibits a Haldane gap, $\Delta_H = 0.41 J_{\text{eff}}=0.023 J$, at $q/\pi = 1/3$ and $q/\pi = 1$. 
The black line in Fig.~\ref{fig:DSF} represents the single magnon band within the reduced Brillouin zone (BZ), fitted using the expression $\omega (q) \approx \Delta_{H} \,\sqrt{1 + \sum_{n=1}^5 a_n\{1 - \cos[n(\pi-3q)]\}}$ with coefficients $a_n$ taken from Ref.~\cite{Affleck2008}. The red line corresponds to the two-magnon dispersion, described by $2\omega(\pi - q/2)$.
The mechanism of single-magnon excitation is shown in Fig. S4(b) of \SM. The lowest excitation corresponds to the breaking of one valence bond solid (VBS) bond, similar to the case of a spin-$1$ chain. Each spin-$1$ is represented as two spin-$1/2$ (shown as two circles in Fig. S4 of \SM), where one spin-$1/2$ on a site forms a singlet dimer with a spin-$1/2$ on a neighboring site, leading to the formation of a VBS state. The lowest excitation involves breaking one spin-$1/2$ dimer, leading to the formation of a triplon which delocalizes throughout the system, as shown in Fig. S4(a) $\&$ (b) of \SM. 

{\it Energy band B.}--- It represents a singleton excitation in one of the trimers in the chain, and this excitation can propagate throughout the chain. The total spin of one single trimer unit changes from $S_{123}=1$ to $S_{123}=0$, which leads to a change in the expectation value of the local bond energy. This quasi-particle costs an energy of $\omega \approx 1.37J$. The weak $q$ dependence of the energy band shows the flat nature of the singleton band, as indicated by $B$ in Fig.~\ref{fig:DSF}. The mechanism of the singleton excitation is shown in Fig. S4(c) of \SM, where one trimer gets excited into $S_{123}=0$ state by breaking two adjacent singlet bonds.

{\it Energy band C and D.}--- The band C and D are degenerate in case of isolated trimer but have different spins $S_{123}=1 $ and $S_{123}=2$. The $S(q,\omega)$ show two split bands C and D and it is generated by the  excitation in one of the trimer of the chain in the gs. In case of band C, one trimer of chain get excited into first excited triplet state as shown in Fig. S4(d) of \SM, but the spin of excited trimer remains same, therefore, small change in local bond energy i.e the gap remains equal to the lowest triplet state excitation energy ($\omega \approx 2J$). This excitation can also propagate through out the chain and have weak $q$-dependence. In band D  spin of one excited trimer goes from $S_{123}=1$ to $S_{123}=2$, as shown in Fig. S4(e) of \SM, which also lead to change in the expectation value of local bond energy. This quasi-particle excitation costs an energy $\omega \approx 2.27J$ for $J'/J = 0.1$, as shown in Fig.~\ref{fig:DSF}.

\begin{figure*}[t]
    \centering
    \includegraphics[width=1.0\linewidth]{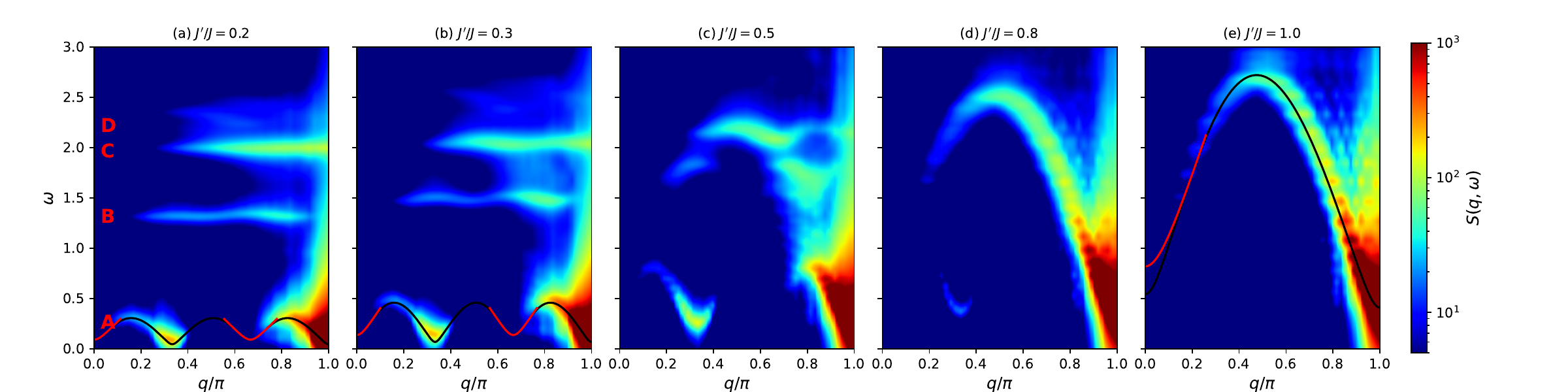}
    \caption{Evolution of the dynamic spin structure factor $S(q, \omega)$ computed using DMRG for a trimer chain of system size $N = 96$ at different coupling ratios $J'/J = 0.2$, $0.3$, $0.5$, $0.8$, and $1.0$, with a broadening factor $\eta = 0.02$. The magnon band is labeled as A, while bands B, C, and D correspond to singleton, triplon, and penton excitations, respectively. For $J'/J = 0.5$ the lower band A is highly dispersive and lies below B band. The C and D bands are appeared to be merged. As the inter-trimer coupling increases to $J'/J = 0.8$,  the B, C, and D bands fully merge into a single dispersive band, while the lower band A remains present. At $J'/J = 1$, the trimers lose their distinct identity, resulting in a magnon continuum. The black line represents the single-magnon dispersion given by $\omega(q)$, while the red line shows the two-magnon dispersion given by $2\omega(\pi-q/2)$.
    }
    \label{fig:evolution}
\end{figure*}

{\it Effect of $J'/J$ variation.}--- We now analyze the effect of $J'/J$ on $S(q, \omega)$ and compare it for five values of $J'/J = 0.1$, $0.3$, $0.5$, $0.8$, and $1$, as shown in Fig.~\ref{fig:evolution}(a–e). In the weak $J'/J$ limit, the low-energy band follows the spectrum of the Hamiltonian in Eq.~\eqref{Eq:isolated}. The lowest band, A, is weakly dispersive, and its bandwidth and the Haldane gap are proportional to $J'/J$ in the perturbative limit. The Haldane gap $\Delta_H$ can be detected by the largest $S(q = \pi, \omega)$ value. The singleton, excited triplon, and penton bands remain well separated up to $J'/J = 0.3$, as shown in Fig.~\ref{fig:evolution}(a) and (b). For $J'/J = 0.5$ the lower band A is highly dispersive and lies below the singleton band. The excited triplon and penton bands are appeared to be merged, as shown in Fig.~\ref{fig:evolution}(c). As the inter-trimer coupling increases to $J'/J = 0.8$,  the singleton, excited triplon, and penton bands fully merge into a single dispersive band, while the lower band A remains present, as shown in Fig.~\ref{fig:evolution}(d). For $J'/J = 1$, the trimers lose their identity and behave as a uniform spin-$1$ chain, exhibiting a gapped excitation at $q = \pi$, as shown in Fig.~\ref{fig:evolution}(e). For this parameter, our calculated single magnon dispersion agrees well with the existing literature \cite{Sharma2025, Affleck2008}.

In the perturbative regime ($J'/J \ll 1$), the energy band B and D increases with $J'$ as the gs energy decreases, while, the C band does not increase much. The major changes in these excitation gaps arise due to variations in the exchange energy (local bond energy) between the excited trimer and its neighboring trimers. The change in local bond energies $\Delta\omega^x(J')$ is approximately written as,
\begin{align}
  \Delta\omega^x(J'/J) & \approx \omega^{x} + 2J_{eff}'' \Big( \langle \psi_{x}| \vec{S}_{i,3} \cdot \vec{S}_{i+1,1} |\psi_{x} \rangle \nonumber \\ &- \langle \psi_{gs}| \vec{S}_{i,3} \cdot \vec{S}_{i+1,1} |\psi_{gs} \rangle \Big)
  \label{Eq:energy_gap}
\end{align}
where $x$ denotes the singleton, triplon, or penton excitation. Here, $\omega^x$ represents the change in energy for an isolated trimer corresponding to a singleton, triplon, or penton excitation. The second term accounts for the local bond energy in the ground state and the excited-state associated with the singleton, triplon, or penton. $J_{eff}''$ is the renormalized effective exchange coupling in perturbative limit, which is approximately equal to $J'/J$. The variation of $\Delta\omega^x(J'/J)$ with $J'$ is depicted in Fig. S5, with a detailed discussion provided in \SM~\cite{suppmat}.

{\it Conclusion and summary.}--- This study provides a comprehensive analysis of the dynamical properties and excitation spectrum of a spin-$1$ trimer chain, highlighting the emergence of various quasi-particles, including magnons, singletons, triplons, pentons, and heptons. In weak inter-trimer coupling regime ($J'/J \ll 1$), the low-energy excitations can be mapped onto an effective spin-$1$ chain consisting of one effective spin-$1$ in each unit cell.  In this limit  lowest magnon excitations has a bandwidth proportional to $J'$ and it is fragments into three smaller windows: $0 \leq q \leq \pi/3$, $\pi/3 \leq q \leq 2\pi/3$, and $2\pi/3 \leq q \leq \pi$, and in the high energy window there are stable high-spin quasi-particles like singleton, excited magnon and penton.  As the ratio approaches $J'/J \simeq 1$  these high spin-quasi particles reduce to conventional magnon excitations of normal spin-1 chain. We also provide a comprehensive understanding of the evolution of these excitation quasi-particles. These excitations can be readily detected using inelastic neutron scattering. This work should motivate the community to synthesize an antiferromagnetic spin-$1$ trimer chain. However, currently, only ferromagnetic spin-$1$ coupled trimer chains have been reported \cite{Masashi2012, Masashi2006, Majumder2015, Yusuf2016, Anup2018}. However, in an ideal ferromagnetically coupled  spin-1 trimer chain, we would expect low energy of spectrum should similar to a ferromagnetically coupled spin-1 chain \cite{Li_2024}, but the high energy spectrum should be similar to that of antiferromagnetically coupled trimers. These findings suggest a rich and complex behavior in the spin-$1$ trimer chain, with potential implications for future studies on spin systems and quantum quasi-particles. These quasi-particles may have potential application in designing spin based devices.

{\it Acknowledgment.}--- We acknowledge very helpful discussions and initial calculations with Sambunath Das, Sumit Haldar. AB and MR acknowledge the financial support from DST-India. We acknowledge National Supercomputing Mission (NSM) for providing computing resources of ‘PARAM RUDRA’ at S.N. Bose National Centre for Basic Sciences, which is implemented by C-DAC and supported by the Ministry of Electronics and Information Technology (MeitY) and Department of Science and Technology (DST), Government of India.

\bibliography{ref_trimer_untracked}
\end{document}


\title{Supplemental Material: Emergent quasi-particles of spin-1 trimer chain}

\author{Manodip Routh}
\thanks{The authors contributed equally to this work.}
\affiliation{Department of Condensed Matter and Materials Physics,
S. N. Bose National Centre for Basic Sciences, Kolkata 700106, India}
\author{Anutosh Biswas}
\thanks{The authors contributed equally to this work.}
\affiliation{Department of Condensed Matter and Materials Physics,
S. N. Bose National Centre for Basic Sciences, Kolkata 700106, India}
\author{Manoranjan Kumar}
\email[Corresponding author: ]{manoranjan.kumar@bose.res.in}
\affiliation{Department of Condensed Matter and Materials Physics,
S. N. Bose National Centre for Basic Sciences, Kolkata 700106, India}

\date{\today}
\maketitle

\noindent\rule{\textwidth}{0.4pt}
We here provide supplemental materials on the following topics in relation to the main text:
\begin{itemize}
    \item[(I)]  Energy spectrum of spin-1 isolated trimer and it's wavefunctions. The wavefunctions are represented in the $S^z$ basis in terms of spin-1 singlet, triplet and free spin. 
    \item[(II)]  First order degenerate perturbation theory for spin-1 trimer chain.
    \item[(III)] Details of numerical method employed in the calculations.
    \item[(IV)] Underlying mechanism of excitations in perturbative limit.
    \item[(V)] Excitation gap in the weak inter-trimer limit ($J'/J \ll 1$). 
\end{itemize}
\noindent\rule{\textwidth}{0.4pt}

\renewcommand{\thesection}{S1}
\section{isolated trimer}
In this subsection we study the magnetic properties and  energetics of the isolated spin-1 trimer and for $J'=0$ the model Hamiltonian is written as
\begin{equation}
    \mathcal{H}=J(\vec{S}_1.\vec{S}_2+\vec{S}_2.\vec{S}_3)
    \label{Eq:isolated_trimer}
\end{equation}
The Hamiltonian can also be written in terms of $\vec{S}_{123}$, $\vec{S}_{13}$ and $\vec{S}_{2}$.

\begin{align}   
\mathcal{H}(\vec{S}_{123}, \vec{S}_{13}, \vec{S}_{2}) &= \frac{J}{2}[(\vec{S}_1+\vec{S}_2+\vec{S}_3)^{2}-\vec{S}_{1}^{2}-\vec{S}_{2}^{2}-\vec{S}_{3}^{2}-2\vec{S}_{1}.\vec{S}_{3}] \nonumber \\
&=\frac{J}{2}[\vec{S}_{123}^{2}-\vec{S}_{1}^{2}-\vec{S}_{2}^{2}-\vec{S}_{3}^{2}-(\vec{S}_1+\vec{S}_3)^{2}-\vec{S}_{1}^{2}-\vec{S}_{3}^{3}] \nonumber\\
&=\frac{J}{2}[\vec{S}_{123}^{2}-\vec{S}_{13}^{2}-\vec{S}_{2}^{2}] 
 \label{eq:eq_ham}
\end{align} 
The  energies of the spin-1 trimer are written as,
\begin{equation}
E(S_{123}, S_{13}, S_{2})=\frac{J}{2}[S_{123}(S_{123}+1)-S_{13}(S_{13}+1)-S_{2}(S_{2}+1)].
 \label{eq:eq_ene}
\end{equation}

For an isolated spin-$1$ trimer, there are $27$ energy states, and Table~\ref{tab:ene_spctr} shows the degeneracies corresponding to each energy level. The wavefunctions of these energy states are provided in Section~\ref{subsec:wavefunction}.

\renewcommand{\thetable}{S1}
\begin{table}[h!]
\centering
     \caption{\centering Energy spectrum of spin-1 trimer chain. }
    \label{tab:ene_spctr}
    \begin{tabular}{|c|c|c|c|c|} \hline 
         $E(S_{123}, S_{13}, S_{2})$&$S_{2}$&  $S_{13}$&  $S_{123}$& Degeneracy\\ \hline 
         $-3J$&  1&  2& 1&3\\ \hline 
          $-2J$& 1&  1& 0&1\\ \hline 
         $-J$&  1&  1& 1 &3\\ \hline 
         $-J$&  1&  2& 2&5\\ \hline 
         $0$&  1&  0& 1&3\\ \hline 
         $J$&  1&  1& 2&5\\ \hline 
         $2J$&  1&  2& 3 &7\\ \hline
 \multicolumn{4}{|c|}{Total number of states}&27\\\hline
    \end{tabular}
\end{table}
\newpage

\renewcommand{\thesubsection}{S1.1}
\subsection{\label{subsec:wavefunction} Wavefunctions for isolated trimer}

\renewcommand{\thesubsubsection}{\textbf{S1.1.1}}
\subsubsection{\textbf{Ground State ($E_0 = -3J$)}}
The ground state (gs) of Hamiltonian in Eq.~\eqref{eq:eq_ham} of a single spin-1 trimer is a three fold degenerate triplet state $S_{123}$ with energy $E_{0}=-3J$. The wavefunctions of the three fold degenerate states $|\psi_M \rangle$, with $M = 1$, $0$, and $-1$, can be written as, 
\begin{align}
   &\ket{\psi_{1}} =\frac{1}{\sqrt{60}}[\ket{-1 1 1}-3\ket{0 0 1}+6\ket{1 -1 1} -3\ket{1 0 0}+\ket{1 1 -1}+2\ket{0 1 0}] ,\\
   &\ket{\psi_{0}} =\frac{1}{\sqrt{60}}[-2\ket{-1 0 1}+3\ket{-1 1 0}+3\ket{0 -1 1} -4\ket{0 0 0} +3\ket{0 1 -1}+3\ket{1 -1 0}-2\ket{1 0 -1}], \\
   &\ket{\psi_{-1}} =\frac{1}{\sqrt{60}}[\ket{-1 -1 1}-3\ket{-1 0 0}+6\ket{-1 1 -1}+2\ket{0 -1 0}- 3\ket{0 0 -1}+\ket{1 -1 -1}].
\end{align}

\renewcommand{\thesubsubsection}{\textbf{S1.1.2}}
\subsubsection{\textbf{Singleton excitation ($E_1 = -2J$)}}

The  1$^{\rm st}$ excited state of a trimer with total spin $S_{123}=0$ has energy $E_1 = -2J$ and it is a non-degenerate state and we call this excitation as singleton. The wavefunction of this trimer state can be written as,
\begin{align}
   &\ket{\psi_{0}} =\frac{1}{\sqrt{6}}[\ket{1 -1 0}-\ket{0 -1 1}-\ket{1 0 -1} +\ket{-1 0 1}+\ket{0 1 -1}-\ket{-1 1 0}] 
\end{align}

\renewcommand{\thesubsubsection}{\textbf{S1.1.3}}
\subsubsection{\textbf{Triplon excitation ($E_2 = -J$)}}

The 2$^{\rm nd}$ excited state with energy $E_2 = - J$ is eight fold degenerate including three fold excited triplet state and five fold degenerate lowest pentate states. The wavefunctions $|\psi_{M}\rangle$ of excited triplet states with total spin $S_{123}=1$ and $M$ represents the multiplicity: 
\begin{align}
   &\ket{\psi_{1}} =\frac{1}{2}[\ket{1 0 0}-\ket{0 0 1}-\ket{1 1 -1}+\ket{-1 1 1}] \\
   &\ket{\psi_{0}} =\frac{1}{2}[\ket{1 -1  0}-\ket{0 -1 1}-\ket{0 1 -1}+\ket{-1 1 0}] \\
   &\ket{\psi_{-1}} =\frac{1}{2}[\ket{1 -1 -1}-\ket{-1 -1 1}-\ket{0 0 -1}+\ket{-1 0 0}]
\end{align}

\renewcommand{\thesubsubsection}{\textbf{S1.1.4}}
\subsubsection{\textbf{Penton excitation ($E_2 = -J$)}}

The five fold pentate states with energy $E_2=-J$ and total spin $S_{123}=2$ can be written as,
\begin{align}
   &\ket{\psi_{2}} =\frac{2}{\sqrt{6}}\ket{1 0 1}-\frac{1}{\sqrt{6}}\ket{1 1 0}-\frac{1}{\sqrt{6}}\ket{0 1 1} \\
   &\ket{\psi_{1}} =\frac{1}{\sqrt{12}}[2\ket{1 -1  1}+\ket{1 0 0}+\ket{0 0 1}-2\ket{0 1 0}-\ket{1 1 -1}-\ket{-1 1 1
   }] \\
   &\ket{\psi_{0}} =\frac{1}{2}[\ket{1 -1 0}+\ket{0 -1 1}-\ket{0 1 -1}-\ket{-1 1 0}] \\
    &\ket{\psi_{-1}} =\frac{1}{\sqrt{12}}[-2\ket{-1 1  -1}-\ket{-1 0 0}-\ket{0 0 -1}-2\ket{0 -1 0}+\ket{-1 -1 1}+\ket{1 -1 -1
   }] \\
   &\ket{\psi_{-2}} =-\frac{2}{\sqrt{6}}\ket{-1 0 -1}+\frac{1}{\sqrt{6}}\ket{-1 -1 0}+\frac{1}{\sqrt{6}}\ket{0 -1 -1} 
\end{align}

\renewcommand{\thesubsubsection}{\textbf{S1.1.5}}
\subsubsection{\textbf{Triplon excitation ($E_3 = 0$)}}

The 3$^{\rm rd}$ excited state with energy $E_3 = 0$ is a three fold degenerate second excited triplet. The wavefunctions for these triplets are written as

\begin{align}
   &\ket{\psi_{1}} =\frac{1}{\sqrt{3}}[\ket{1 1 -1}-\ket{0 1 0}+\ket{-1 1 1}] \\
   &\ket{\psi_{0}} =\frac{1}{\sqrt{3}}[\ket{1 0 -1}-\ket{0 0 0}+\ket{-1 0 1}] \\
   &\ket{\psi_{-1}} =\frac{1}{\sqrt{3}}[\ket{1 -1 -1}-\ket{0 -1 0}+\ket{-1 -1 1}]
\end{align}

\renewcommand{\thesubsubsection}{\textbf{S1.1.6}}
\subsubsection{\textbf{Penton excitation ($E_4 = J$)}} 

The 4$^{\rm th}$ excited state with energy $E_4 = J$ is a five fold degenerate pentate states. The wavefunctions for these five states can be expressed as:

\begin{align}
   &\ket{\psi_{2}} =\frac{1}{\sqrt{2}}[\ket{1 1 0}-\ket{0 1 1}] \\
   &\ket{\psi_{1}} =\frac{1}{2}[\ket{1 0  0}-\ket{0 0 1}+\ket{1 1 -1}-\ket{-1 1 1
   }] \\
   &\ket{\psi_{0}} =\frac{1}{\sqrt{12}}[\ket{1 -1 0}-\ket{-1 1 0}+\ket{0 1 -1}-\ket{-0 -1 1}]+\frac{1}{3}[\ket{1 0 -1}-\ket{-1 0 1}] \\
    &\ket{\psi_{-1}} =\frac{1}{2}[-\ket{-1 0  0}+\ket{0 0 -1}-\ket{-1 -1 1}+\ket{1 -1 -1
   }] \\
    &\ket{\psi_{-2}} =\frac{1}{\sqrt{2}}[-\ket{-1 -1 0}+\ket{0 -1 -1}] 
\end{align}

\renewcommand{\thesubsubsection}{\textbf{S1.1.7}}
\subsubsection{\textbf{Hepton excitation ($E_4 = 2J$)}}  

The highest excited state with energy $E_4 = 2J$ is a seven fold degenerate heptate state. The wavefunctions of these states can be expressed as:

\begin{align}
   &\ket{\psi_{3}} =\ket{1 1 1} \\
   &\ket{\psi_{2}} =\frac{1}{\sqrt{3}}[\ket{1 0  1}+\ket{1 1 0}+\ket{0 1 1}]\\
   &\ket{\psi_{1}} =\frac{1}{\sqrt{15}}[\ket{1 -1 1}+2\ket{1 0 0}+2\ket{0 0 1}+2\ket{0 1 0}+\ket{1 1 -1}+\ket{-1 1 1}]\\
    &\ket{\psi_{0}} =\frac{1}{\sqrt{10}}[\ket{1 -1  0}+\ket{0 -1 1}+\ket{1 0 -1}+2\ket{0 0 0}+\ket{-101}+\ket{01-1}+\ket{-110}] \\
   &\ket{\psi_{-1}} =\frac{1}{\sqrt{15}}[\ket{-11 -1}+2\ket{-1 0 0}+2\ket{0 0 -1}+2\ket{0 -1 0}+\ket{-1 -1 1}+\ket{1 -1 -1}]\\
    &\ket{\psi_{-2}} =\frac{1}{\sqrt{3}}[\ket{-1 0 - 1}+\ket{-1 -1 0}+\ket{0 -1 -1}]\\ 
    &\ket{\psi_{-3}} =\ket{-1 -1 -1} 
\end{align}

\newpage
\renewcommand{\thesection}{S2}
\section{Perturbation Theory of spin-1 trimer chain}

In order to understand the nature of the spin-$1$ trimer chain in the weak inter-trimer exchange ($J'/J \ll 1$) limit we perform the 1$^{st}$ order degenerate perturbation theory. Each spin-1 trimer behaves as spin-1 site and the trimer chain exhibits translational invariance with a unit cell consisting of three spins and maintains rotational symmetry. Consequently, the effective Hamiltonian must preserve both translation symmetry and SU(2) symmetry. The total Hamiltonian of two trimer unit can be expressed as:
\begin{align}
\mathcal{H}_{tot}&=2\mathcal{H}_{0}+\mathcal{H'}\nonumber \\
     &=2\mathcal{H}_{0}+J'\vec{S}_{i,3}.\vec{S}_{i+1,1}
     \label{Eq:eff_ham}
\end{align} 
where $\mathcal{H}_0$ is the Hamiltonian for a isolated trimer in Eq.~\eqref{Eq:isolated_trimer}, $\vec{S}_{i,3}$ and $\vec{S}_{i+1,1}$ are the $3^{rd}$ spin operator of $i$-th trimer unit and $1^{st}$ spin operator of $(i+1)$-th trimer unit, respectively. At $J'=0$, the ground state of the Hamiltonian in Eq.~\eqref{Eq:eff_ham} can be written as product states of two isolated spin-1 trimer units. Therefore, the new basis are $\ket{1_i 1_{i+1}}$, $\ket{1_i 0_{i+1}}$, $\ket{1_i -1_{i+1}}$, $\ket{0_i 1_{i+1}}$, $\ket{0_i 0_{i+1}}$, $\ket{0_i -1_{i+1}}$, $\ket{-1_i 1_{i+1}}$, $\ket{-1_i 0_{i+1}}$, $\ket{-1_i -1_{i+1}}$, where $\ket{1}$, $\ket{0}$, $\ket{-1}$ are the ground state wavefunctions for each isolated trimer unit. 

The perturbative part of the Hamiltonian in Eq.~\eqref{Eq:eff_ham} can be written in terms of creation and annihilation operators as,
\begin{equation}
   \mathcal{H}'= J' \vec{S}_{i,3}.\vec{S}_{i+1,1}=\frac{J'}{2}({S}_{i,3}^{+}{S}_{{i+1},1}^{-}+{S}_{i,3}^{-}{S}_{{i+1},1}^{+})+ J' {S}_{i,3}^{z}{S}_{{i+1},1}^{z}
   \label{Eq:Hprime}
\end{equation}

The nonzero matrix elements of the operator  $S_{i,3/1}^{z}$ in the new basis are $\bra{1_{i}} S_{i,3/1}^{z} \ket{1_{i}} = \frac{3}{4}$, $\bra{-1_{i}} S_{i,3/1}^{z} \ket{-1_{i}} = -\frac{3}{4}$. The complete set of matrix elements is provided in Table~\ref{tab:spin_operators}(a). Similarly, the nonzero matrix elements of the operator ${S_{i,3/1}^{+}}$ in the new basis are $\bra{1_i}{S_{i,3/1}^{+}}\ket{0_i}=\frac{3}{2\sqrt{2}}$ and $\bra{0_i}{S_{i,3/1}^{+}}\ket{-1_i}=\frac{3}{2\sqrt{2}}$. The complete set of matrix elements is provided in Table~\ref{tab:spin_operators}(b). Similarly, the nonzero matrix elements of the operator ${S_{i,3/1}^{-}}$ are $\bra{0_i}{S_{i,3/1}^{-}}\ket{1_i}=\frac{3}{2\sqrt{2}}$ and $\bra{-1_i}{S_{i,3/1}^{+}}\ket{0_i}=\frac{3}{2\sqrt{2}}$.The complete set of matrix elements is provided in Table~\ref{tab:spin_operators}(c). 

\renewcommand{\thetable}{S2}
\begin{table}[h!]
    \centering
    \caption{Matrix elements of spin operators in the new basis.}
    \label{tab:spin_operators}
    \begin{subtable}{0.32\textwidth}
        \centering
        \caption{Matrix elements for $S^{z}$.}
        \label{tab:sz_op}
        \begin{tabular}{|c|c|c|c|}
            \hline
            ${S}_{i,3/1}^{z}$ & $\ket{1_i}$ & $\ket{0_i}$ & $\ket{-1_i}$ \\ \hline
            $\ket{1_i}$      & $\frac{3}{4}$ & 0   & 0   \\ \hline
            $\ket{0_i}$      & 0   & 0   & 0   \\ \hline
            $\ket{-1_i}$     & 0   & 0   & $-\frac{3}{4}$ \\ \hline
        \end{tabular}
    \end{subtable}
    \hfill
    \begin{subtable}{0.32\textwidth}
        \centering
        \caption{Matrix elements for $S^{+}$.}
        \label{tab:splus_op}
        \begin{tabular}{|c|c|c|c|}
            \hline
            ${S}_{i,3/1}^{+}$ & $\ket{1_i}$ & $\ket{0_i}$ & $\ket{-1_i}$ \\ \hline
            $\ket{1_i}$      & 0   & $\frac{3}{2\sqrt{2}}$ & 0   \\ \hline
            $\ket{0_i}$      & 0   & 0   & $\frac{3}{2\sqrt{2}}$ \\ \hline
            $\ket{-1_i}$     & 0   & 0   & 0   \\ \hline
        \end{tabular}
    \end{subtable}
    \hfill
    \begin{subtable}{0.32\textwidth}
        \centering
        \caption{Matrix elements for $S^{-}$.}
        \label{tab:sminus_op}
        \begin{tabular}{|c|c|c|c|}
            \hline
            ${S}_{i,3/1}^{-}$ & $\ket{1_i}$ & $\ket{0_i}$ & $\ket{-1_i}$ \\ \hline
            $\ket{1_i}$      & 0   & 0   & 0   \\ \hline
            $\ket{0_i}$      & $\frac{3}{2\sqrt{2}}$ & 0   & 0   \\ \hline
            $\ket{-1_i}$     & 0   & $\frac{3}{2\sqrt{2}}$ & 0   \\ \hline
        \end{tabular}
    \end{subtable}
\end{table}

Using these matrices, the perturbative Hamiltonian in Eq.~\eqref{Eq:Hprime} can be expressed as,

\begin{align}
{\mathcal{H}^{'}_{i,i+1}}&=\frac{1}{2}[\begin{pmatrix}
0 & \frac{3}{2\sqrt{2}} & 0 \\
0 & 0 & \frac{3}{2\sqrt{2}} \\
0 & 0 & 0
\end{pmatrix}
\otimes
\begin{pmatrix}
0 & 0 & 0 \\
\frac{3}{2\sqrt{2}} & 0 & 0 \\
0 & \frac{3}{2\sqrt{2}} & 0
\end{pmatrix}+\begin{pmatrix}
0 & 0 & 0 \\
\frac{3}{2\sqrt{2}} & 0 & 0 \\
0 & \frac{3}{2\sqrt{2}} & 0
\end{pmatrix}
\otimes
\begin{pmatrix}
0 & \frac{3}{2\sqrt{2}} & 0 \\
0 & 0 & \frac{3}{2\sqrt{2}} \\
0 & 0 & 0
\end{pmatrix}
] \nonumber\\
&\quad+ \begin{pmatrix}
\frac{3}{4} & 0 & 0 \\
0 & 0 & 0 \\
0 & 0 & -\frac{3}{4}
\end{pmatrix}
\otimes
\begin{pmatrix}
\frac{3}{4} & 0 & 0 \\
0 & 0 & 0 \\
0 & 0 & -\frac{3}{4}
\end{pmatrix}
 \label{Eq:Hprime_matrix}
\end{align}

Next, we introduce a new set of spin operators in the effective spin-$1$ basis to express $\mathcal{H}'$. Here, $\tau^{+}$, $\tau^{-}$, and $\tau^{z}$ denote the creation, annihilation, and $z$-component of the spin operators, respectively, for the effective spin-$1$.

\begin{center}  
$\tau_{i}^{+}\ket{-1_i}=\sqrt{2}\ket{0_i}$, $\tau_{i}^{+}\ket{0_i}=\sqrt{2}\ket{1_i}$,
$\tau_{i}^{+}\ket{1_i}=0$
\end{center}
\begin{center}
$\tau_{i}^{-}\ket{-1_i}=0$, $\tau_{i}^{-}\ket{0_i}=\sqrt{2}\ket{-1_i}$, $\tau_{i}^{-}\ket{1_i}=\sqrt{2}\ket{0_i}$
\end{center}
\begin{center}
$\tau_{i}^{z}\ket{-1_i}=-\ket{-1_i}$, $\tau_{i}^{z}\ket{0_i}=0$, $\tau_{i}^{z}\ket{1_i}=\ket{1_i}$
\end{center}

Rearranging the terms in Eq.~\eqref{Eq:Hprime_matrix} we get,
\begin{align}
{\mathcal{H}_{i,i+1}^{'}}&=\frac{9}{16}\times\{\frac{1}{2}[\begin{pmatrix}
0 & \sqrt{2} & 0 \\
0 & 0 & \sqrt{2} \\
0 & 0 & 0
\end{pmatrix}
\otimes
\begin{pmatrix}
0 & 0 & 0 \\
\sqrt{2} & 0 & 0 \\
0 & \sqrt{2} & 0
\end{pmatrix}+\begin{pmatrix}
0 & 0 & 0 \\
\sqrt{2} & 0 & 0 \\
0 & \sqrt{2} & 0
\end{pmatrix}
\otimes
\begin{pmatrix}
0 & \sqrt{2} & 0 \\
0 & 0 & \sqrt{2} \\
0 & 0 & 0
\end{pmatrix}
] \nonumber\\
&\quad+ \begin{pmatrix}
1 & 0 & 0 \\
0 & 0 & 0 \\
0 & 0 & -1
\end{pmatrix}
\otimes
\begin{pmatrix}
1 & 0 & 0 \\
0 & 0 & 0 \\
0 & 0 & -1
\end{pmatrix}\}
\nonumber\\
&\quad=\frac{9}{16}\times\{\frac{1}{2}(\tau_{i}^{+}\otimes\tau_{{i+1}}^{-}+\tau_{i}^{-}\otimes\tau_{{i+1}}^{+})+\tau_{i}^{z}\otimes\tau_{{i+1}}^{z}\}
 \label{Eq:Eq13}
\end{align}

The final form of the perturbative Hamiltonian is given by  
\begin{equation}  
    \mathcal{H}_{\text{eff}} = J_{\text{eff}} \sum_{i} {\tau}_{i} \cdot {\tau}_{i+1},  
\end{equation}  
where $J_{\text{eff}} = \frac{9J'}{16J}$. Table~\ref{tab:compare_nrg} presents a comparison of the energy obtained from the two-site perturbation approach and the energy of two trimers for different values of $J'/J$.

\renewcommand{\thetable}{S3}
\begin{table}[h]
    \centering
    \caption{Energy comparison between the energy of two site perturbation and energy of two trimers in different values of $J'/J$.}
    \label{tab:compare_nrg}
    \begin{tabular}{|c|c|c|} \hline 
         $J'/J$&  Energy of two trimers& Energy using two site perturbation theory\\ \hline 
         0.01&  -6.0113& -6.0113\\ \hline 
         0.05&  -6.0572& -6.0563\\ \hline 
         0.10&  -6.1161& -6.1125\\ \hline 
         0.15&  -6.1767& -6.1688\\\hline
    \end{tabular}
    
\end{table}

\renewcommand{\thesection}{S3}
\section{Numerical Method}

{\it DMRG.}--- We apply the density-matrix renormalization group (DMRG) technique to calculate the dynamics for the large system size of the spin-1 trimer chain. The DMRG is a state-of-the-art numerical technique to calculate accurate ground states and a few low-lying excited energy states of strongly interacting quantum systems \cite{srwhite1992,srwhite1993}. It is based on the systematic truncation of irrelevant degrees of freedom. We use DMRG in both open and periodic boundary conditions. For PBC we apply a modified DMRG algorithm \cite{Dey_Maiti_Kumar_2016} which avoids multiple times renormalization of operators in constructing the superblock and this algorithm provides excellent accuracy for the one-dimensional systems. For OBC we include additional spin-$1/2$ at the edge of the chain to avoid the edge effect \cite{Affleck2008, Sharma2025}. In our DMRG calculations, we consider the trimer chain with $N = 96$ spins and retain up to $1000$ density-matrix eigenstates $m$ during the renormalization process. Retaining the large number of $m$ ensures the high accuracy of our results, with the largest discarded weight being maintained at an impressively low level of approximately $10^{-10}$.

\vspace{0.5cm}

{\it Dynamical structure factor.}--- The dynamical structure factor (DSF) is defined as:
\begin{equation}
S(q,\omega) = \sum_{n} \frac {\left| \langle \psi_n | S_q^{\alpha} | \psi_0 \rangle \right|^2}{(E_n - (E_0 + \omega) + i\eta)},
\label{Eq:SQW}
\end{equation}
where, $E_0$ and $E_n$ are the energies of the gs and the $n^{th}$ excited state, respectively. The $\omega$, $\eta$ and $q$ represent the energy, broadening factor and momentum, respectively. $|\psi_0\rangle$ is the ground state wavefunction and $|\psi_n\rangle$ is the $n^{th}$ excited state wavefunction. if $\alpha$ denotes the $x$, $y$, and $z$ components of spin, we define 
\begin{equation}
    S_q^{\alpha} = \sqrt{\frac{2\pi}{N}} \sum_{j} e^{iqj} S_j^{\alpha}.
\end{equation}

DSF $S(q,\omega)$ is finite only for selected states and the selection rule for a finite $S(q, \omega)$ can be written as follows: (a) $\langle \psi (S'^z) | S^z(q) | \psi (S^z) \rangle \neq 0$ for $S^z = S'^z$, (b) $\langle \psi (S'^z) | S^\pm(q) | \psi (S^z) \rangle \neq 0$ for $S^z = S'^z \pm 1$, and for the total spin ($S_T$), (c) $\langle S_T | S^\mu(q) | S_{T'} \rangle = 0$ if $\left| S_T - S_{T'} \right| \neq 0, 1$ or if $S_T = S_{T'} = 0$. Therefore, we focus on selected excitations that can be probed using inelastic neutron scattering.

The DSF of the system is calculated using the combination of the DMRG and the correction vector method which is a  well-established numerical approach \cite{Soos1989, Ramasesha1995, Ramasesha1997, Till1999, Jeckelmann2002, Nocera2016}. In this work, we use both conjugate gradient \cite{Soos1989, Ramasesha1995, Ramasesha1997, Till1999} and Krylov space approach \cite{Nocera2016}. However, all the simulations here are done using Krylov space approach.

\renewcommand{\thesubsection}{S3.1}
\subsection{Finite-size effect of the dynamical structure factor $S(q, \omega)$}

Figure~\ref{fig:dsf_compare_N}(a-b) presents a comparison of the dynamical structure factor $S(q, \omega)$ at $J'/J = 1$ for a trimer chain with system sizes $N = 72$ and $96$, using a broadening factor of $\eta = 0.05$. The finite-size effects are found to be minimal. The black line in Fig.~\ref{fig:dsf_compare_N}(a-b) represents the single magnon band within the reduced Brillouin zone (BZ), fitted using the expression $\omega (q) \approx \Delta_{H} \,\sqrt{1 + \sum_{n=1}^5 a_n\{1 - \cos[n(\pi-q)]\}}$ with coefficients $a_n$ taken from Ref.~\cite{Affleck2008}. The red line corresponds to the two-magnon dispersion, described by $2\omega(\pi - q/2)$.

\renewcommand{\thefigure}{S3.1}
\begin{figure}[h]
    \centering
    \includegraphics[width=\linewidth]{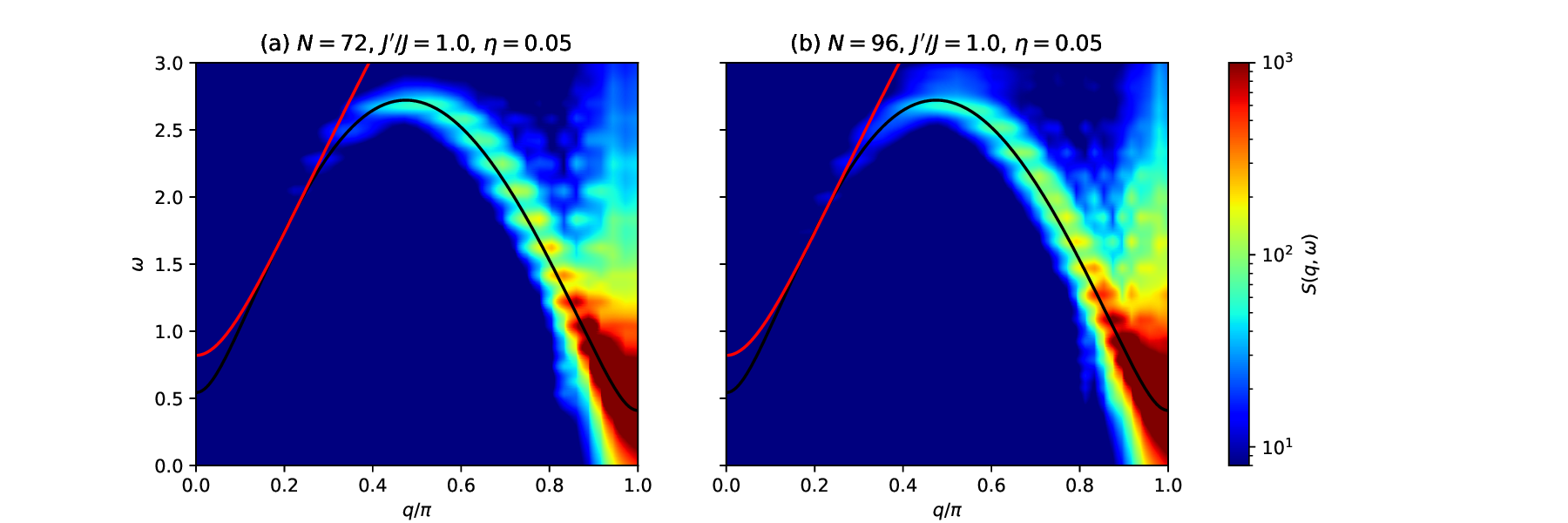}
    \caption{Dynamical structure factor $S(q, \omega)$ for a trimer spin chain at $J'/J = 1.0$ is shown for system sizes (a) $N = 72$ and (b) $N = 96$. A fixed broadening factor of $\eta = 0.05$ is used in both cases. The results exhibit minimal dependence on system size.
    }
    \label{fig:dsf_compare_N}
\end{figure}

We also present the dynamical structure factor for various values of $m$, where $m$ denotes the number of states retained in the density matrix of the system block during the renormalization step in DMRG. Figure~\ref{fig:dsf_compare_max}(a–d) shows $S(q, \omega)$ at $J'/J = 1$ for a system size of $N = 96$ and different values of $m$. We find that $m > 500$ is required to accurately capture the DSF.

\renewcommand{\thefigure}{S3.2}
\begin{figure}[h]
    \centering
    \includegraphics[width=\linewidth]{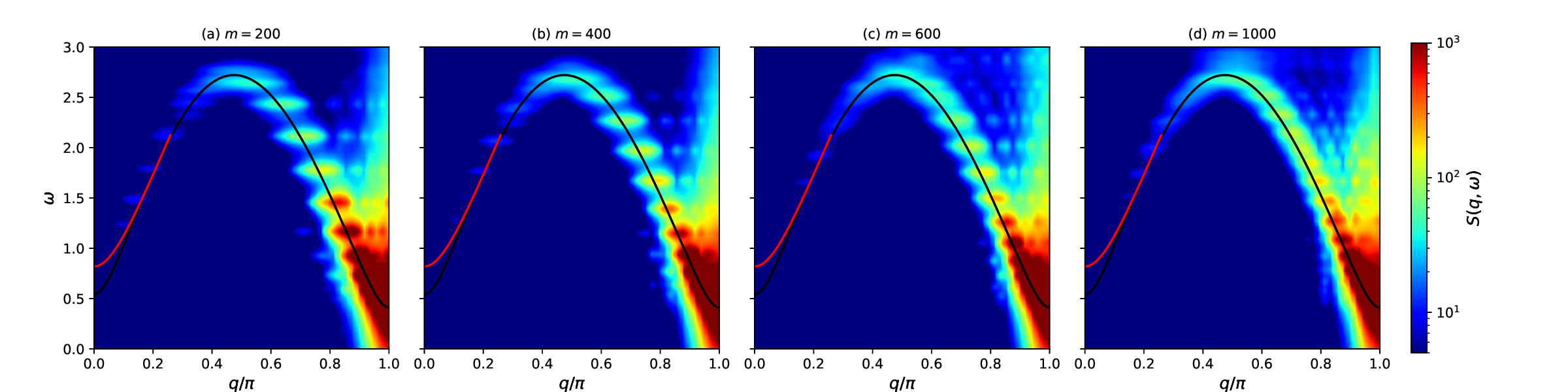}
    \caption{Dynamical structure factor $S(q, \omega)$ for the trimer spin chain at $J'/J = 1.0$ shown for different numbers of states ($m$) retained in the density matrix during the DMRG calculation. The system size is fixed at $N = 96$, and the broadening factor is set to $\eta = 0.05$ in all cases.}
    \label{fig:dsf_compare_max}
\end{figure}

\renewcommand{\thesection}{S4}
\section{Mechanism of excitations in perturbative limit}

The schematic representation of the ground state and its possible excitations is illustrated in Fig.~\ref{fig:vbs}. In the ground state, each site consists of a effective spin-$1$ entity, represented by a solid rectangle, an open circle inside denoting an internal effective spin-$1/2$ degree of freedom. Two spin-$1/2$ from the neighboring sites form a singlet state or valance bond state ~\cite{aklt1987,schollwock2002}, indicated by a solid ellipse connecting two adjacent spin-$1/2$. Excitations arise from breaking a singlet bond which equivalent to breaking a inter-trimer bond and generating magnon excitations as shown in Fig.~\ref{fig:vbs} (a).  A singleton excitation occurs when one of  trimers gets excited into $S=0$ state of the trimer and it can move anywhere in the chain and resulting the renormalizing the bond energy in the system as shown in the Fig.~\ref{fig:vbs} (b). Similarly, a excited triplon excitation emerges in one of the trimers and these can move anywhere in chain and resulting  renormalization of bond energy as well. The mechanism is shown in Fig.~\ref{fig:vbs} (c). Similarly, a penton or heptate quasi-particles are generated when one of the triplet spin-$1$ trimer goes to $S=2$ or $S=3$ trimer state and moves in the chain leading to change in local bond energy as shown in the Fig.~\ref{fig:vbs} (d) and e respectively. The intensity of single magnon excitations is the highest, followed by triplon excitations, which have a higher intensity than singleton and penton excitations.

\renewcommand{\thefigure}{S4}
\begin{figure}[h]
    \centering
     \includegraphics[width=0.5\linewidth]{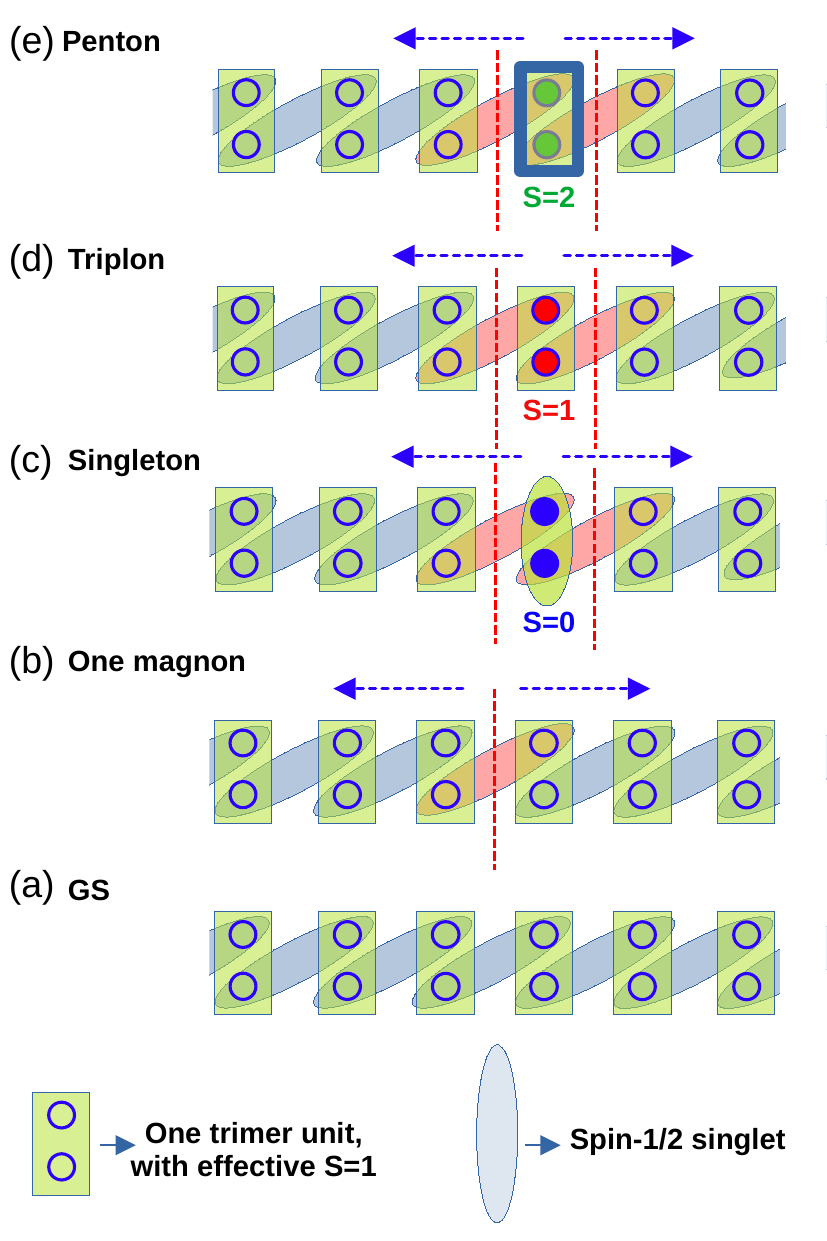}
    \caption{Schematic representation of the ground state and its possible excitations. (a) The ground state configuration, where each site with effective spin-$1$ is represented by a solid rectangle, and an open circle inside denotes a single effective spin-$1/2$. A solid ellipse indicates a spin-$1/2$ singlet formed between two neighboring spin-$1/2$. (b) A magnon excitation generates from breaking an inter-trimer bond. (c) A singleton excitation arises from breaking two adjacent bonds associated with a singlet ($S=0$). Likewise, (d) a triplon excitation and (e) a penton excitation emerge from breaking two neighboring bonds associated with total spin $S=1$ and $S=2$, respectively.}
    \label{fig:vbs}
\end{figure}

\renewcommand{\thesection}{S5}
\section{Excitation gap in the weak inter-trimer limit $J'/J \ll 1$}
The ground state (gs) energy of a spin$-1$ trimer chain with periodic boundary condition (PBC) in the perturbative limit can be written as,
\begin{align}
    E^{gs} &= \sum_{i=1}^{N/3} E_{i}^{gs}+  \sum_{i=1}^{N/3} J_{eff} \langle \psi_{gs}| \vec{S}_{i,3} \cdot \vec{S}_{i+1,1} |\psi_{gs} \rangle \nonumber \\
           &= \frac{N}{3} E_{i}^{gs}+\frac{N}{3}J_{eff} \langle \psi_{gs}| \vec{S}_{i,3} \cdot \vec{S}_{i+1,1} |\psi_{gs} \rangle 
\end{align}
where, $N$ is the total number spins, $E^{gs}_{i}$ is the energy of a single trimer with $S_{123}=1$, $J_{eff} = \frac{9J'}{16J}$ is the effective exchange in the perturbative limit and $\langle \psi_{gs}|\vec{S}_{i,3} \cdot \vec{S}_{i+1,1} |\psi_{gs} \rangle$ is the expectation value of local bond energy in the gs.

The excited state energy can be written approximately as,
 \begin{align}
    E^{x} &\approx \sum_{i=1}^{N/3-1} E_{i}^{gs}+E_{i}^{x}+ J_{eff} \sum_{i=1}^{N/3-2}\langle \psi_{gs}| \vec{S}_{i,3} \cdot \vec{S}_{i+1,1} |\psi_{gs} \rangle + 2J_{eff}' \langle \psi_{x}| \vec{S}_{i,3} \cdot \vec{S}_{i+1,1} |\psi_{x} \rangle
\end{align}
where $x$ denotes the singleton, triplon, or penton excitation. The first term represents the ground state energy of $(N/3-1)$ isolated trimers, the second term corresponds to the excitation energy of an isolated trimer (with $x$ denoting a singleton, triplon, or penton), and the third term consists of the local bond energy in the ground state for $(N/3-2)$ bonds, along with the excited state energy of the two adjacent bonds associated with the excited singleton, triplon, or penton. Here, $J_{\text{eff}}'$ denotes the effective exchange interaction for the two adjacent bonds associated with the excited singleton, triplon, or penton. \\

For an isolated trimer, the singleton, excited triplon, and penton excitations occur at $\omega = J$ and $2J$, respectively, and the excited triplon and penton states are degenerate. However, this degeneracy is lifted for a coupled trimer when $J' > 0$. We find that the singleton and penton energy gap increase with $J'$ significantly, whereas the excited triplon energy gap does not increase very much. The major changes in these excitation gaps arise due to variations in the exchange energy (local bond energy) between the excited trimer and its neighboring trimers. In the chain, each trimer is connected to two neighboring trimers; therefore, the approximate  energies $\omega^x(J')$ is written as,
\begin{align}
  \Delta\omega^x(J') & = E^{x}-E^{gs} \nonumber \\
  & \approx (E_{i}^{x}-E_{i}^{gs}) + 2J_{eff}' \langle \psi_x| \vec{S}_{i,3} \cdot \vec{S}_{i+1,1} |\psi_x \rangle - 2J_{eff}\langle \psi_{gs}| \vec{S}_{i,3} \cdot \vec{S}_{i+1,1} |\psi_{gs} \rangle \nonumber \\
  & \approx \omega^{x} + 2J_{eff}'' \left( \langle \psi_x| \vec{S}_{i,3} \cdot \vec{S}_{i+1,1} |\psi_x \rangle - \langle \psi_{gs}| \vec{S}_{i,3} \cdot \vec{S}_{i+1,1} |\psi_{gs} \rangle \right)
  \label{Eq:energy_gap}
\end{align}

Here, $\omega^x$ represents the change in energy for an isolated trimer corresponding to a singleton, triplon, or penton excitation. $J_{eff}''$ is the renormalized effective exchange coupling in perturbative limit, which is approximately equal to $J'$. Fig.~\ref{fig:ene_gap} shows the linear variation of $\Delta\omega^{x} (J')$ as a function of $J'$. The solid circles indicate numerical values of $\Delta\omega^{x} (J')$ computed using ED for a system size of $N=6$ trimer sites, while the open circles correspond to $\Delta\omega^{x} (J') - \omega^x$ calculated from Eq.~\eqref{Eq:energy_gap}. The computed values from Eq.~\eqref{Eq:energy_gap} agrees well with the numerical results of $S(q, \omega)$.

\renewcommand{\thefigure}{S5}
\begin{figure}[h]
\centering
    \includegraphics[width=0.60\linewidth]{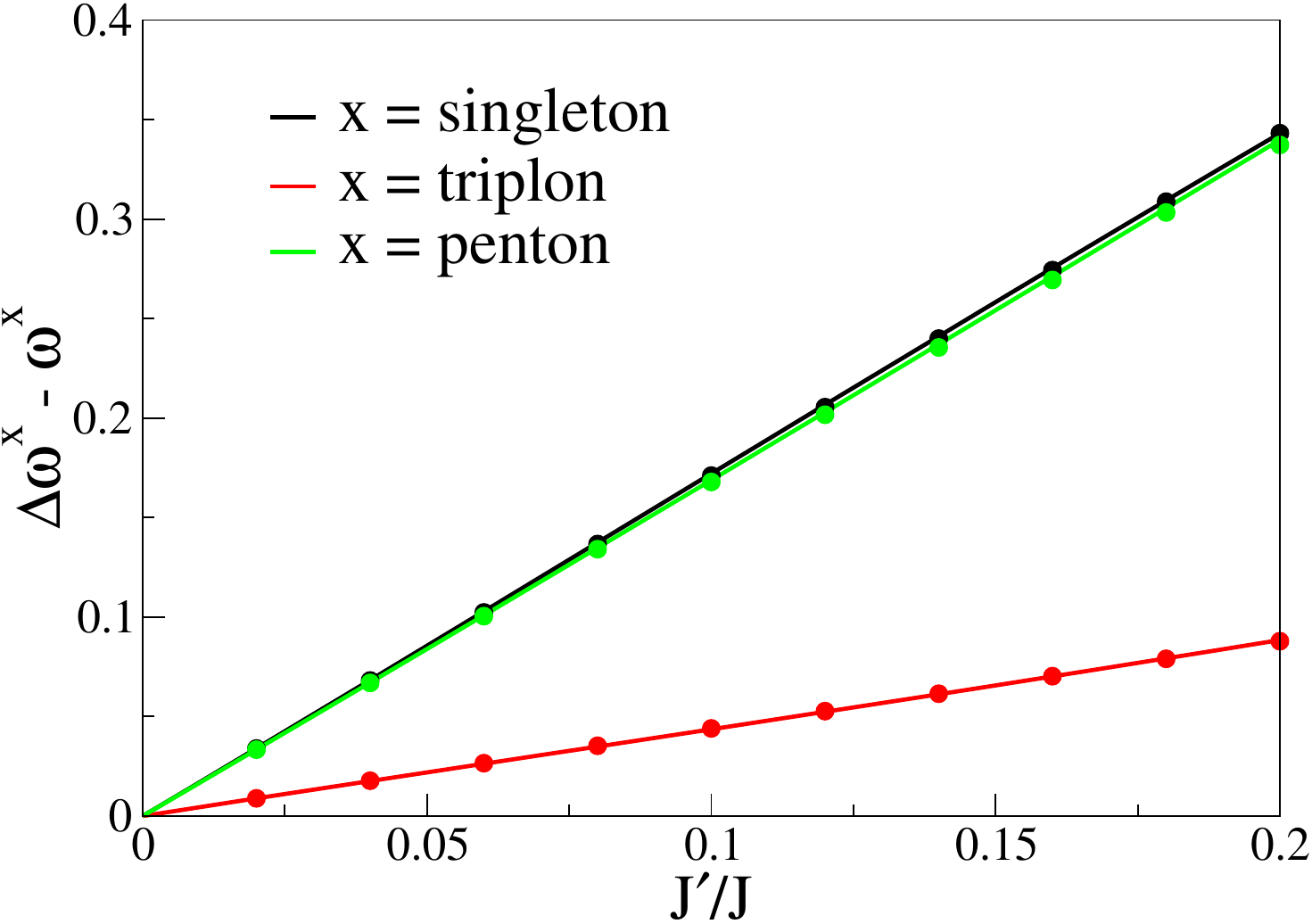}
    \caption{Variation of the energy gap, $\Delta\omega^{x} (J') - \omega^x$, for the singleton, triplon, and penton as a function of $J'/J$ for an $N = 6$ site system in the weak inter-trimer coupling regime ($J'/J \leq 0.2$). The solid line represents the computed values from Eq.~\eqref{Eq:energy_gap}, while the solid circles denote the numerically obtained energy gap from the DSF [Eq.~\eqref{Eq:SQW}].}
    \label{fig:ene_gap}
\end{figure}

\newpage
\bibliography{ref_trimer_untracked}